\documentclass[]{spie}

\usepackage{graphicx}
\usepackage{amssymb}

\title{Resonant light-light interaction in slab waveguides: \\ angular filters and spectral hole burning}

\author{Andrey V. Gorbach\supit{a}, Victor Fleurov\supit{b}, Sergej Flach\supit{a}  and Andrey E. Miroshnichenko\supit{c}
\skiplinehalf
\supit{a}Max-Planck-Institut f\"ur Physik komplexer Systeme,\\ N\"othnitzerstr. 38, Dresden 01187, Germany \\
\supit{b}Beverly and Raymond Sackler Faculty of Exact Sciences,
School of Physics and Astronomy,\\ Tel Aviv University, Tel Aviv 69978, Israel\\
\supit{c}Nonlinear Physics Centre,
Research School of Physical Sciences and Engineering,\\
Australian National University, Canberra ACT 0200, Australia
}

\authorinfo{Further author information: (Send correspondence to A.V.G.)\\A.V.G.: E-mail: gorbach@pks.mpg.de}

\begin{document}
\maketitle

\begin{abstract}
We consider the process of low-power light scattering by optical solitons in a slab waveguide with homogeneous and inhomogeneous refractive index core.
We observe resonant reflection (Fano resonance) as well as resonant
transmission of light by optical solitons at certain incident angles. The resonance position can be controlled experimentally by changing the soliton intensity and the
relative frequency detuning between the soliton and the probe light beams.
\end{abstract}


\section{Introduction}
\label{sec_intro}
Recently the problem of plane wave scattering by various time-periodic potentials
has attracted much attention, as it has been shown that some interesting effects
such as resonant reflection of waves \cite{CAF98,KS98,KK01} can be observed when
dealing with non-stationary scattering centers. These resonant reflection effects were demonstrated to be similar to the well-known Fano resonance \cite{Fan61} for some nonlinear systems \cite{FMF03,FMF+03}.
In one-dimensional (1D) systems they can lead to a \emph{total resonant reflection}
of waves. The time-periodic scattering centers can originate from the nonlinearity
in a spatially homogeneous system \cite{Aub97,FW98,leshouches,CFK04}. The principle
underlying idea of this phenomenon is that a time-periodic scattering
potential induces several harmonics. In general, these harmonics can be either
inside or outside the plane wave spectrum, 'open' and 'closed' channels, respectively \cite{FMF03,FMF+03}. 
The presence of closed channels is equivalent to a local
increase of the spatial dimensionality, i.e. to an appearance of alternative paths
for the plane wave propagation. This can lead to novel interference effects, such as
the resonant reflection of waves.

The total resonant reflection of plane waves can
be similarly arranged by means of a static scattering center, provided the system dimensionality has been artificially locally increased (e.g. in composite materials)
\cite{KKL+03,KLK+05,Kos97,GSY+02}. However, such static configurations have at least
two significant disadvantages. Firstly, they do not provide any flexibility in tuning
the resonance parameters: the resonant values of plane wave parameters are fixed by
the specific geometry. Secondly, it may be a nontrivial technological task to
introduce locally additional degrees of freedom for plane waves in the otherwise 1D 
system. Time-periodic scattering potentials appear to be much more promising: they
can be relatively easily generated (e.g. laser beams, microwave radiation, localized 
soliton-like excitations), and they provide us with an opportunity to tune the
resonance, since all the resonance parameters depend on the parameters of the
potential (e.g. amplitude, frequency) and thus can be 'dynamically' controlled by
some parameter, e.g. the laser beam intensity.

Recently we demonstrated \cite{FFG+05} a possibility of experimental observation of the 
Fano resonance in the scattering process of light by optical spatial solitons. The 
proposed setup is based on the slab optical waveguide with a specially designed inhomogeneous slab structure. The Fano resonance manifests itself through a resonant 
reflection of a weak-power probe beam from the spatial soliton, formed in the slab, at
a certain incident angle, defined by the soliton power. We also conjectured that 
the Fano resonance could be observed in a purely homogeneous slab waveguide, 
provided a frequency detuning between the interacting light beams is
introduced. In this paper we perform a detailed analysis of the interaction process
of detuned beams in the slab waveguide. We confirm our previous suggestions as for
the possibility of a Fano resonance observation in the homogeneous waveguide core. We
demonstrate, that the resonance position is strongly depending both on the value of
frequency detuning and the material characteristics -- the frequency dependencies of
its linear and non-linear susceptibility tensors' components. A combination of frequency 
detuning and modulation of the core refractive index provides us with an additional 
flexibility in controlling the resonance position. This might be an essential
advantage regarding the possibility of engineering of tunable optical filters.
We also study the process of wavepacket scattering by optical solitons and demonstrate a
possibility of Fano resonance detection through an observation of holes in the
spectrum of transmitted light.

The paper is organized as follows. In Sec.~\ref{sec_setup} we introduce the setup and 
develop the corresponding model, describing two beams interaction in the slab waveguide.
In Sec.~\ref{sec_transm} we compute the transmission coefficient for plane waves, 
scattered by the soliton, in different waveguide configurations: 
a purely homogeneous slab with a space independent refractive index, and an inhomogeneous slab with
specially designed stepwise modulation of the core refractive index -- triple-well configuration. In both configurations 
the Fano resonance -- vanishing
of the transmission coefficient at a certain spectral parameter
of the plane wave -- can be observed. In Sec.~\ref{sec_wp} we discuss the process of
wave-packet scattering by the soliton. The effect of spectral hole burning occurs 
as the consequence of the Fano resonance. Finally, in Sec.~\ref{sec_concl} we conclude.

\section{The Setup and model equations}
\label{sec_setup}
\begin{figure}
\begin{tabular}[b]{cc}
a)\includegraphics[angle=270, width=0.5\textwidth]{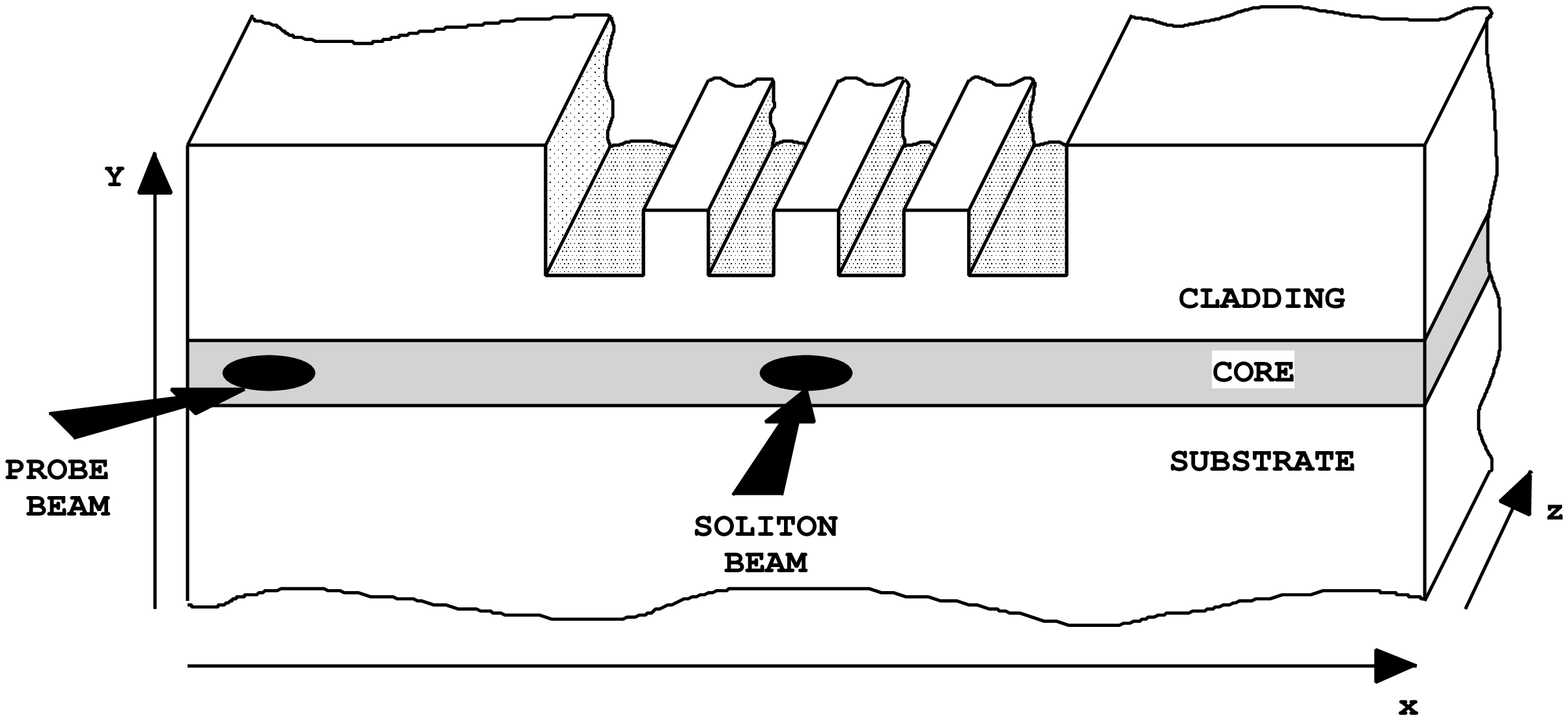} &
b)\includegraphics[angle=270, width=0.4\textwidth]{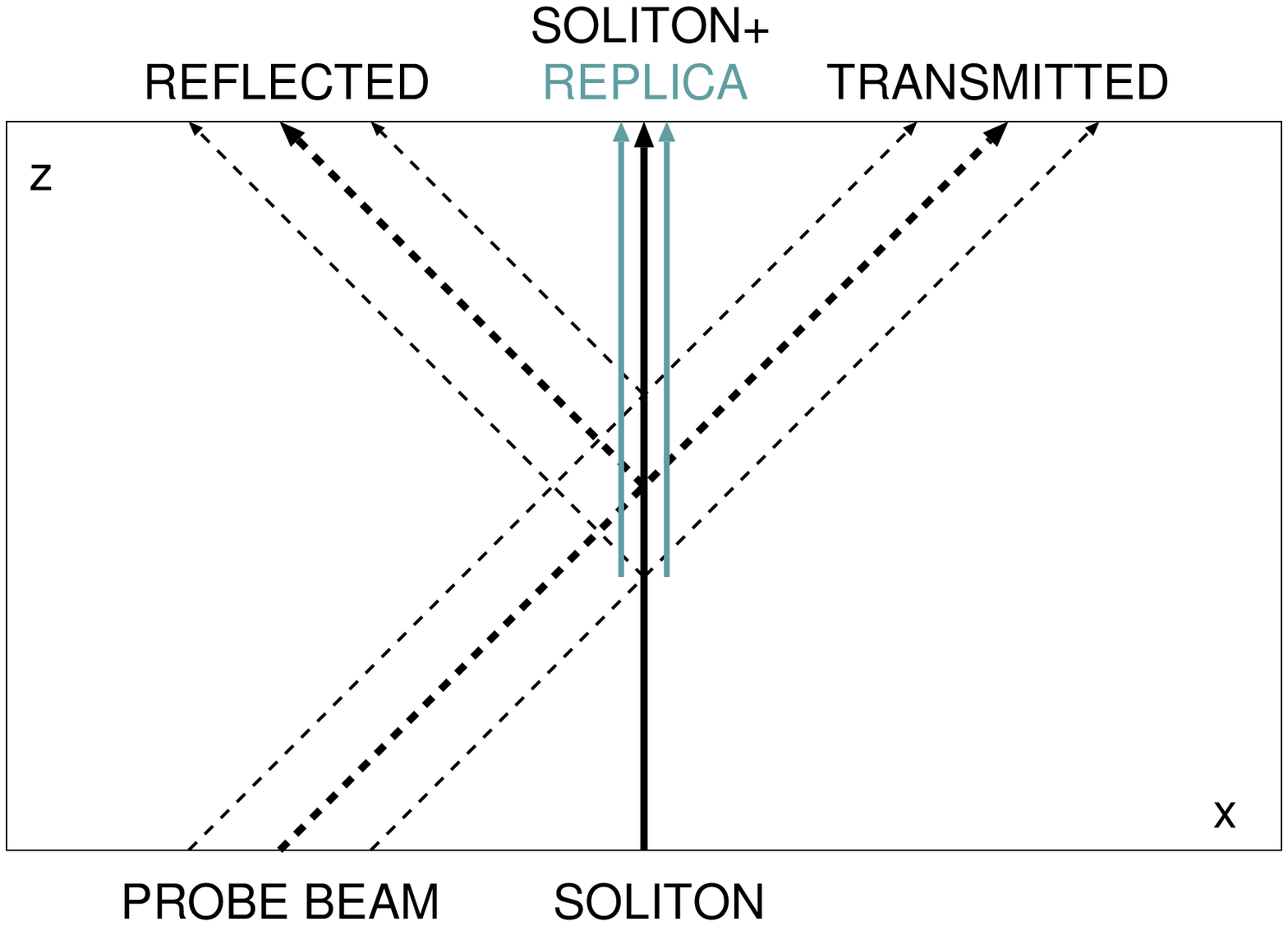}
\end{tabular}
\caption{(a) Schematic structure of the experimental setup.
The soliton beam is sent along the $z$-axis, while the probe
beam propagates in the $xz$-plane at some angle to the soliton. 
(b) Scheme of the scattering process in the slab (see the main text for details).}
\label{fig:waveguide}
\end{figure}

We consider a weak-power probe beam interacting with a spatial optical soliton in the 
planar (slab) waveguide. The soliton is generated in the waveguide by a laser beam
injected into the slab along the $z$-direction, see Fig.~\ref{fig:waveguide}.
The soliton beam light is confined in the $y$-direction (inside the slab) by the
total internal reflection. The localization of light in the $x$-direction (the spatial soliton propagates along the $z$-direction) is ensured by the balance between linear 
diffraction and an instantaneous Kerr-type nonlinearity. The probe beam is sent at some 
angle to the soliton. It has small enough power so that the Kerr nonlinearity of the
medium is negligible outside the soliton core. It is important, that the process of
light scattering by a spatial soliton is assumed to be \emph{stationary} in time,
i.e. the light is quasi-monochromatic. The analogy with the above time-periodic 
scattering problems comes from the possibility to interpret the spatial propagation
along the $z$-direction as time \cite{ZS72,kivshar}. Thus the angle, at which the
probe beam is sent to the $z$-axis plays the role of a parameter similar to
the frequency (or wave number) of plane waves in 1D systems.


Light propagation inside the slab is governed by the Helmholtz equation \cite{kivshar,akhmediev}
\begin{equation}
{\mathcal {\vec{r}}} \cdot \left({\mathcal{\vec{r}}} \cdot 
\vec{\mathcal {E}} \right)- {\mathcal {\vec{r}}}^2 \vec{\mathcal {E}}
=-\frac{1}{c^2}\frac{\partial^2 \vec{\mathcal{E}}}{\partial t^2} -
\frac{4\pi}{c^2}\frac{\partial^2 \vec{\mathcal{P}}^{L}}{\partial t^2}-\frac{4\pi}{c^2}\frac{\partial^2 \vec{\mathcal{P}}^{NL}}{\partial t^2}\;,
\label{helmholtz}
\end{equation}
where $\vec{\mathcal{E}}(\vec{r},t)$ is the electric component of the field, 
$\vec{\mathcal{P}}^{L}(\vec{r},t)$ and $\vec{\mathcal{P}}^{NL}(\vec{r},t)$
are linear and nonlinear components of the polarization vector, respectively. They
are connected with the electric field amplitude through the linear and nonlinear susceptibility tensors
\begin{eqnarray}
\label{pl}
\vec{{\mathcal P}}^L \left(\vec{r},t \right)&=&
\int_{-\infty}^{\infty}dt_1 \chi^{(1)}(\vec{r},t-t_1) \vec{{\mathcal E}} 
\left(\vec{r},t_1 \right)\;,\\
\vec{{\mathcal P}}^{NL} \left(\vec{r},t \right)&=&\int\int\int_{-\infty}^{\infty}dt_1dt_2dt_3
\chi^{(3)}(\vec{r},t-t_1,t-t_2,t-t_3) \vec{{\mathcal E}} 
\left(\vec{r},t_1\right)\vec{{\mathcal E}} \left(\vec{r},t_2\right)
\vec{{\mathcal E}} \left(\vec{r},t_3\right)\;.
\end{eqnarray}
where the medium response is considered to be local in space.
The waveguide is supposed to be homogeneous along the $z$-direction, so that the
susceptibility tensors do not depend on the $z$ coordinate. The slab waveguide geometry 
assumes a slight stepwise order of 1\% relative modulation of these tensors across the $y$ direction throughout the sample. 
In addition we allow for a local modulation of the susceptibility tensors in the $x$ direction in the finite region $|x|<L_s$
(the "scattering region"), where the soliton is located, see
Fig.~\ref{fig:waveguide}(a). This modulation results in a stepwise modulation of the
linear refractive index $n=n(x,y)$.

We take the scalar approach in what follows, assuming that the electric components
of the field in both beams are polarized along the $y$-direction (TM mode):
$\vec{\mathcal{E}}(\vec{r},t)=\hat{y}\mathcal{E}(\vec{r},t)$, $\hat{y}$ is the unit 
vector along the $y$-axis. Within this approach we neglect other polarization
components, arising from the first term in l.h.s. of Eq.~(\ref{helmholtz}) in the
case of a spatially inhomogeneous refractive index of the medium. This can be
done under the assumption that both linear (due to the waveguide specific geometry)
and nonlinear (due to the Kerr effect) modulations of the refractive index are
small \cite{akhmediev} (weak-guidance approximation).

Further simplification is achieved by considering the $y$-dependence of the field
to be quasi-uniform all over the system: $\mathcal{E}(\vec{r},t) = \mathcal{B}(y)
\cdot\mathcal{E}_{\bot}(x,z,t)$ which is in fact true for an ideal single-mode
slab waveguide. Any modulation of the refractive index within the slab in the $x$ or
$z$ directions generally destroys this uniform distribution. A stepwise change of
the refractive index leads to an excitation of radiative modes in the vicinity of the
step. The associated rate of losses is proportional to the refractive index variation 
which is of the order of 1\% in our case. Since such small losses don't play any 
significant role in the resonant effects we describe here, we may neglect them in what
follows.

Finally, we assume that both the probe and the soliton beams are quasi-monochromatic
but propagating at different frequencies $\omega_0$ and $\omega_0+\delta$, respectively. It is straightforward to show that the nonlinear interaction of the two beams results
in a replica beam at the frequency $\omega_0-\delta$. Therefore, the following
three beam ansatz
\begin{equation}
\mathcal{E}_{\bot}(x,z,t)=\frac12\left\{S(x,z)\exp[-i\omega_0 t] + 
L(x,z)\exp[-i(\omega_0+\delta) t]+R(x,z)\exp[-i(\omega_0-\delta) t]+c.c.\right\}
\label{three_beam}
\end{equation}
can be applied. Here the functions $S(x,z)$, $L(x,z)$ and $R(x,z)$ correspond to the soliton, probe beam and replica beam envelopes, respectively; $\omega_0$ is the
carrier frequency of the soliton beam. Using the ansatz~(\ref{three_beam}) together
with the above approximations the set of equations
\begin{eqnarray}
\label{sol_eq}
\frac{\partial^2 S}{\partial z^2}+ \frac{\partial^2 S}{\partial x^2} + n_0^2(x) S+
\alpha_0 \left|S\right|^2 S =0\;,\\
\label{pb_eq}
\frac{\partial^2 L}{\partial z^2}+ \frac{\partial^2 L}{\partial x^2}
+n_+^2(x)
L+\alpha_+\left[2|S|^2L+S^2 R^*\right]=0\;,\\
\label{rb_eq}
\frac{\partial^2 R}{\partial z^2}+ \frac{\partial^2 R}{\partial x^2}
+n_-^2(x)R+
\alpha_-\left[2|S|^2R+S^2 L^*\right]=0\;,
\end{eqnarray}
for the beam envelopes is derived from the Helmholtz equation~(\ref{helmholtz}).
Here $n_0(x)\equiv n(x,\omega_0)$ and $\alpha_0\equiv \alpha(\omega_0)$ are the
effective linear refractive index in the slab and the nonlinear Kerr coefficient
for the soliton beam at the frequency $\omega_0$ with the functions $n(x,\omega)$ and
$\alpha(\omega)$ given by
\begin{eqnarray}
\label{nw}
n^2(x,\omega)&=&1+4\hat{\chi}_{yy}^{(1)}(x,\omega)-\lambda^2(\omega)\;,\\
\alpha(\omega)&=&3\pi\hat{\chi}_{yyyy}^{(3)}(\omega)/2\;,
\label{aw}
\end{eqnarray}
$\lambda\sim d\mathcal{B}/dy$ is the slab waveguide mode constant;
$\hat{\chi}^{(i)}$ are the Fourier components of the corresponding susceptibility
tensors (we assume them to be real, so that all possible losses in the waveguide
are neglected), $n_{\pm}(x)$ and $\alpha_{\pm}$ correspond to the effective
refractive indeces and Kerr coefficients for the probe and replica beams, propagating
at different frequencies:
\begin{eqnarray}
\label{npm}
n_{\pm}^2(x)&=&(1\pm\delta/\omega_0)^2 \cdot n^2(x,\omega_0\pm \delta)\;,\\
\label{apm}
\alpha_{\pm}&=&(1\pm\delta/\omega_0)^2\cdot \alpha(\omega_0\pm \delta)\;.
\end{eqnarray}
The dimensionless spatial units $(\omega_{0}/c)\cdot(x,z)\rightarrow (x,z)$ are used.
All terms nonlinear in $L$ and $R$ were omitted in 
Eqs.~(\ref{sol_eq}), (\ref{pb_eq}), and (\ref{rb_eq}) due to the smallness of the probe and replica beam amplitudes. Thus, treating these beams as small perturbations to
the soliton solution, we neglect their possible influence on the soliton shape and
properties. Outside the region $|x|<L_s$ the refractive indices
$n_{\pm}(|x|>L_s)\equiv n_{\pm,0}$ and $n_{0}(|x|>L_s)\equiv n_{0,0}$ don't
vary in space.

The spatial optical soliton represents a special class of solutions of
Eq.~(\ref{sol_eq}) in the form $S(x,z)=C(x)\exp(i\beta z)$ with $C(x)$ describing
the exponentially localized profile of the soliton, $C(x)|_{x\rightarrow \pm 
\infty}\rightarrow 0$. $\beta$ is the only soliton parameter (the $z$-component
of its wave-vector), $\beta>n_{0,0} $. The soliton profile function $C(x)$
satisfies the stationary nonlinear Schr\"odinger equation (NLS)
\begin{equation}
\label{shrod}
\frac{d^2 C}{d x^2}
+ \left[n_0^2(x)-\beta^2\right] C+
\alpha_0 \left|C\right|^2 C =0,
\end{equation}

Once the soliton solution is found from Eq.~(\ref{shrod}), we substitute it into 
Eqs.~(\ref{pb_eq},\ref{rb_eq}) and obtain the two equations
\begin{eqnarray}
\label{l+OB}
\frac{\partial^2 L}{\partial z^2}+ \frac{\partial^2 L}{\partial x^2}
+n_+^2(x)
L+\alpha_+\left[2|C|^2L+C^2 R^*\exp(i2\beta
z)\right]=0\;,\\
\label{r+OB}
\frac{\partial^2 R}{\partial z^2}+ \frac{\partial^2 R}{\partial x^2}
+n_-^2(x)
R+\alpha_-\left[2|C|^2R+ C^2 L^*\exp(i2\beta
z)\right]=0\;.
\end{eqnarray}
for the probe and replica beams, coupled to each other through the soliton.
The soliton acts as an external potential for the two small-amplitude beams and
consists of two different parts. Treating the $z$ coordinate of light propagation
as an effective time we coin them as 'DC' ($\sim |C|^2$) and 'AC' ($\sim C^2$) parts.
While the former is the soliton induced nonlinear refractive index modulation,
the latter results from the {\em coherent} interaction between the beams.

The soliton part, $C(x)\rightarrow 0$, is negligible outside the scattering
region $|x|<L_s$ and the refractive indices do not vary there. Then
Eqs.~(\ref{l+OB}) and (\ref{r+OB}) reduce to simple wave equations for
plane waves $L(x,z),R(x,z)\sim \exp(ik_x x + i k_z z)$ with the relation
between the $x$- and $z$- components of the wave vector (different for the two beams)
\begin{equation}
\label{spectr}
k_x^2+k_z^2=n_{\pm,0}^2\;,
\end{equation}
being analogous to the dispersion relation $\omega(k)$ of plane waves in 1D 
time-dependent systems. This implies, that far away from the soliton the allowed
$k_z$ values for the freely propagating probe and replica beams are bounded by the intervals $|k_z|\le n_{+,0}$ and $|k_z|\le n_{-,0}\;$, respectively.

The soliton acts as a harmonic generator (due to the AC part of the potential), so
that the coupled probe and replica beams have two different propagation constants
along the $z$-direction, and the general solution of Eqs.~(\ref{l+OB}) and (\ref{r+OB})
are looked for in the form
\begin{eqnarray}
\label{2-channel-1}
L(x,z)&=&A(x)\exp[ik_z z]\;,\\
\label{2-channel-2}
R(x,z)&=&B(x)\exp[i(2\beta-k_z)z]\;.
\end{eqnarray}
We choose $0 \le k_z \le n_{+,0}\;$, so that it satisfies the dispersion
relation (\ref{spectr}) with a {\em real} $k_x$ value and the probe beam wave
can freely propagate far away from the scattering region. Depending on the $\beta$
and $n_{-,0}\;$ values the propagation constant $(2\beta-k_z)$ of the replica beam
can be either above or below the $n_{-,0}\;$ value. In the latter case both the probe
and the replica beams will be present in the transmitted and reflected light.
This corresponds to the so-called multi-channel scattering process. Here we will,
however, focus on a simpler situation of the single-channel scattering process,
when only the probe beam can freely propagate outside the scattering
region, while the propagation constant of the replica beam is above the $n_{-,0}\;$
value and the replica beam is exponentially localized around the scattering region.
Hence we have one open ($A$) and one closed ($B$) channel in this case. The
corresponding scattering process is schematically shown in Fig.~\ref{fig:waveguide}(b). 
Such a single channel scattering process occurs, in particular, when the frequency dependence of the refractive index $n(x,\omega)$ is negligible in the 
$[\omega_0-\delta,\omega_0+\delta]$ frequency interval.

Note, that in the limit $\delta\rightarrow 0$ both the probe and the soliton beams have the same frequency. 
However, the concept of two channels persists, since the two components $A$ and $B$ with different propagation constants are excited due to the beam interaction \cite{FFG+05}.

Substituting the ansatz~(\ref{2-channel-1}) and (\ref{2-channel-2}) into (\ref{l+OB}), (\ref{r+OB}) the
equations
\begin{eqnarray}
\label{2ch-1}
A^{\prime\prime}=\left[k_z^2-n_+^2(x)\right]A-\alpha_+\left[
2|C|^2 A+ C^2 B^*
\right],\\
B^{\prime\prime}=\left[(2\beta-k_z)^2-n_-^2(x)\right]B-
\alpha_-\left[
2|C|^2 B+ C^2 A^*  \right].
\label{2ch-2}
\end{eqnarray}
are obtained for the two channel amplitudes $A$ and $B$ coupled to each
other via the AC part of the soliton scattering potential
\cite{FMF03,FMF+03,FFG+05}. By removing the coupling between the channels, 
i.e. dropping the term $\sim C^2$, the equation for the
$A$ ($B$) channel becomes equivalent to the stationary Schr\"odinger equation for
an effective particle with the energy $E_A=n_{+,0}^2-k_z^2$ 
($E_B=n_{-,0}^2-(2\beta-k_z)^2$) in the external potential
\begin{equation}
\label{veff}
 V_{eff}(x)=[n_{\pm,0}^2-n^2(x)]-
2\alpha_{\pm}|C(x)|^2
\end{equation}
which is the sum of 'geometrical' and 'soliton' parts. The spectra of these two
effective systems consist of continuous parts, corresponding to the dispersion
relation (\ref{spectr}), and discrete sets of localized levels due to the
external potential $V_{eff}$. If the conditions of a single-channel scattering
hold, the continuum parts of these two spectra do not intersect. However, a localized
level(s) of the closed channel may enter the continuum part of the open channel
spectrum. This corresponds to the Fano resonance condition~\cite{Fan61}. Thus,
we expect to observe a resonance between an extended open channel state and the 
bound state of the closed channel as long as there is a weak inter-channel
coupling. In that limit the resonance
position is close to the position of the localized level of the closed channel
inside the open channel continuum band (obtained for uncoupled channels) \cite{FMF03,FMF+03}. Increasing the inter-channel coupling, the resonance position
may shift and even completely disappear. We use the weak inter-channel
coupling limit as the starting point when searching for the resonance and then follow
it while increasing the AC potential strength to its proper value given in Eqs.~(\ref{2ch-1}) and (\ref{2ch-2}).

\section{Transmission coefficient for plane waves}
\label{sec_transm}
To compute the plane wave transmission coefficient $T(k_x)$ we solve Eqs.~(\ref{2ch-1}) and (\ref{2ch-2}) with the boundary conditions \cite{FMF03,FMF+03}
$A(x\rightarrow +\infty)=\tau \exp(ik_x x)$, $A(x\rightarrow -\infty)=\exp(ik_x
x) + \rho \exp(-ik_x x)$, and $B(x\rightarrow +\infty)=F\exp(-\kappa x)$,
$B(x\rightarrow -\infty)=D\exp(\kappa x)$. The amplitudes of transmitted, $\tau$,
and reflected, $\rho$, waves in the open channel define the transmission and
reflection coefficients $T=|\tau|^2=1-|\rho|^2=1-R$, respectively. Amplitudes $F$
and $D$ describe spatially exponentially decaying closed channel excitations
with the inverse localization length
$\kappa=[4\beta^2-4\beta\sqrt{n_{-,0}^2-k_x^2}-k_x^2]^{0.5}$.

We start with the simplest case of a homogeneous slab waveguide
$n_{0}(x)\equiv n_{0,0}\;,\; n_{\pm}(x)\equiv n_{\pm,0}\;$. In this case
$C(x)$ is the well-known stationary NLS soliton \cite{ZS72}
\begin{equation}
\label{nls_soliton}
C(x)=\sqrt{\frac{2\left(\beta^2-n_{0,0}^2\right)}{\alpha_0}}\frac{1}{\cosh\left[
\sqrt{\beta^2-n_{0,0}^2}\cdot x\right]}\;.
\end{equation}
The soliton-induced effective potential~(\ref{veff}) has several localized
levels\cite{Landafshiz} in the spectrum of the uncoupled closed channel with the propagation constants
$k_z^{(j)}$
\begin{equation}
k_z^{(j)}=2\beta-\left\{ n_{-,0}^2 + \frac{\alpha_{-}}{\alpha_{0}} \frac{\beta^2-n_{0,0}^2}{4} \left[\sqrt{1 + 16\frac{\alpha_-}{\alpha_0}}
- (2j-1)\right]^2 \right\}^{0.5}, \qquad j=1,2,...n\;.
\label{loc_levels}
\end{equation}
where the number $n$ of levels is determined by the condition
\begin{equation}
n<\frac12\left(-1+\sqrt{1+16\frac{\alpha_-}{\alpha_0}}\right)\;.
\label{num_levels}
\end{equation}

\begin{figure}
\begin{tabular}{cc}
\includegraphics[angle=270, width=0.45\textwidth]{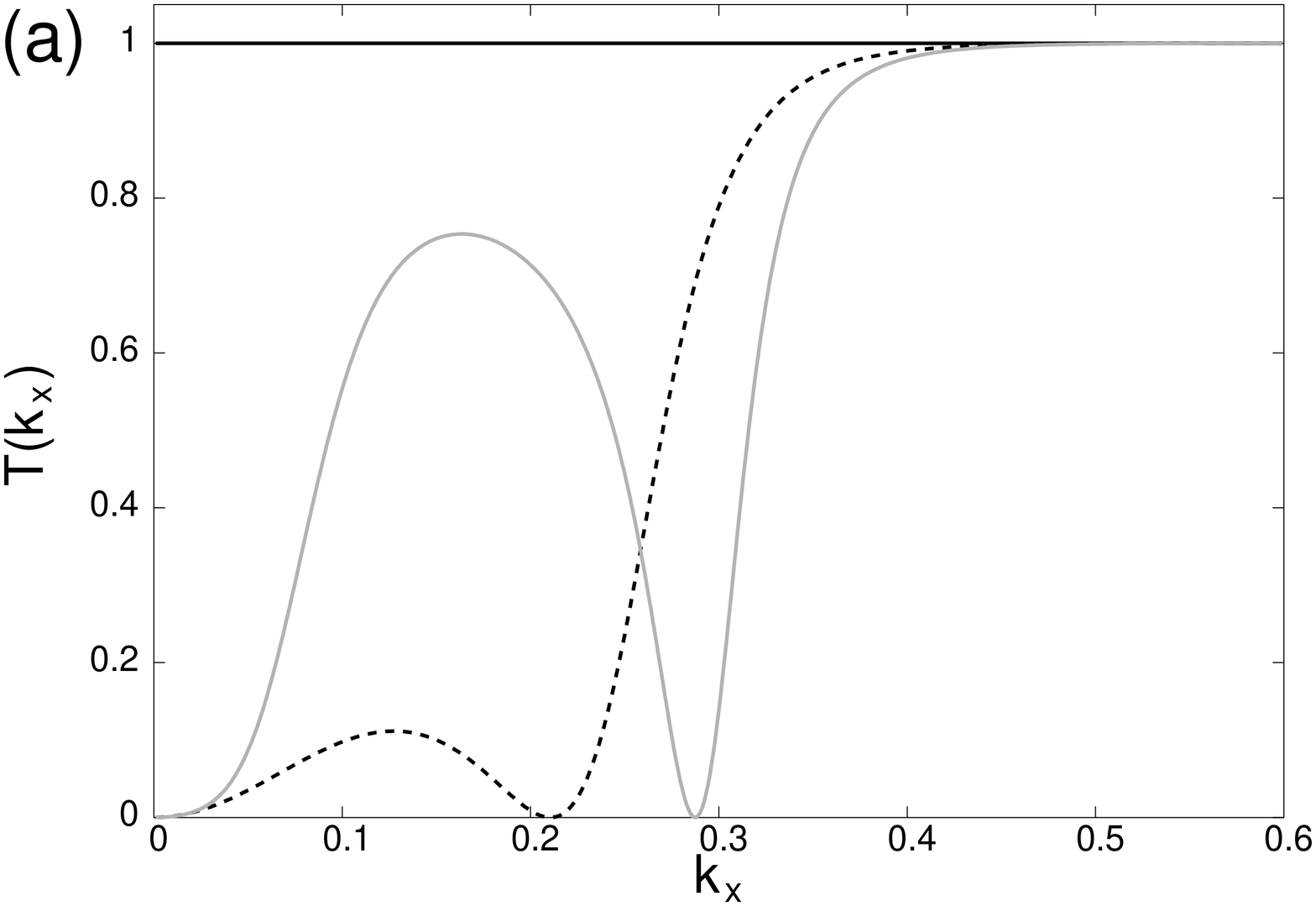}&
\includegraphics[angle=270, width=0.45\textwidth]{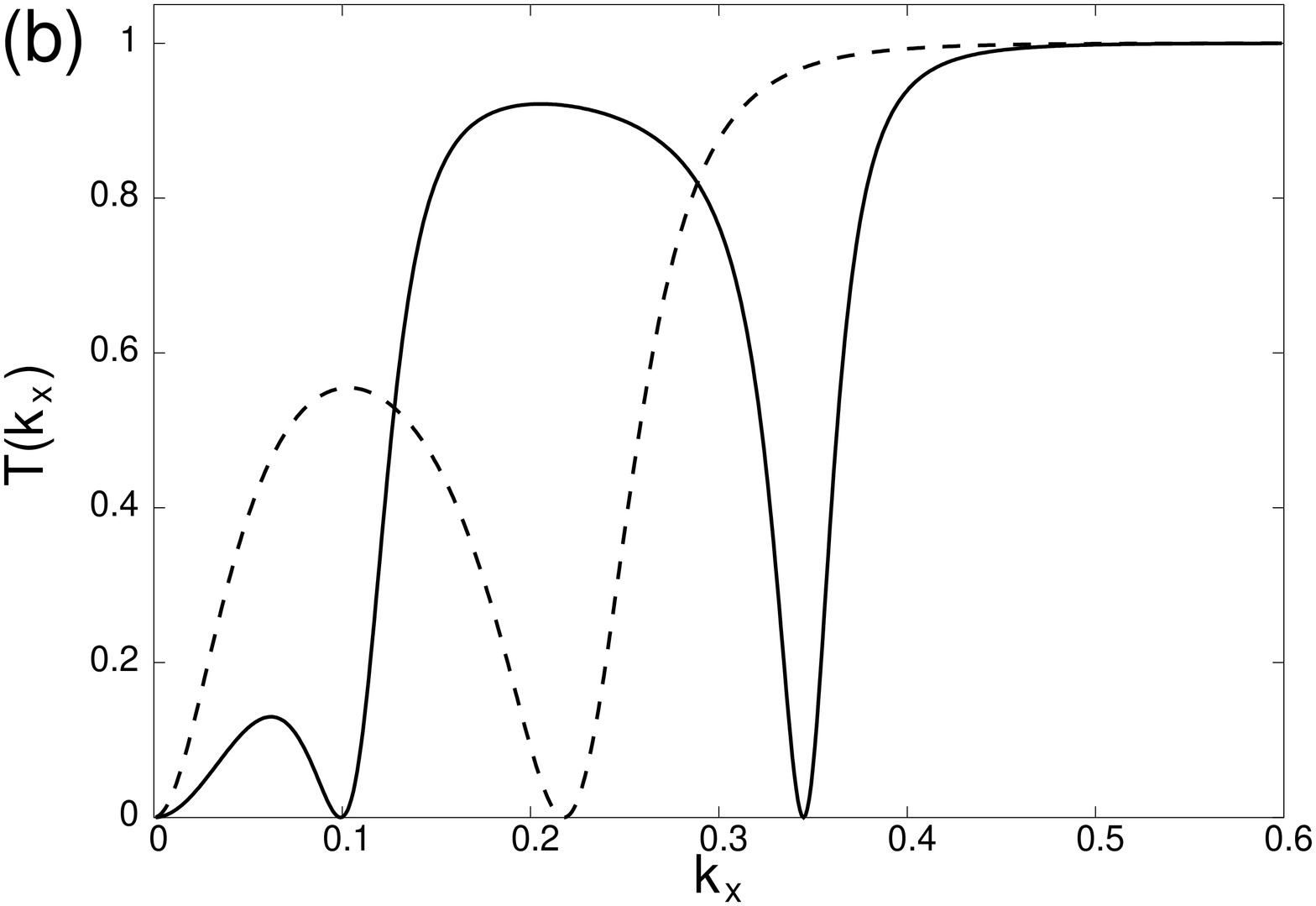}
\end{tabular}
\caption{Transmission coefficient $T(k_x)$ of the detuned probe beam through the
optical soliton with $\beta\approx 1.4743$, $\beta^2-n_{0,0}^2=0.1$,
in the case of a homogeneous slab waveguide with $n_{0,0}=1.44$, $\alpha_0=1.0$.
Parameters $n_{\pm,0}$ and $\alpha_{\pm}$ are determined by
Eqs.~(\ref{npms},\ref{apms}) with different values of $\Delta_n$ and $\Delta_\alpha$: (a) $\Delta_n=\Delta_\alpha=0$ (solid black curve),
 $\Delta_n=\Delta_\alpha=-0.3$ (dashed curve), $\Delta_n=\Delta_\alpha=-0.5$ (solid gray curve); (b) $\Delta_n=-0.5$, $\Delta_\alpha=-0.6$ (solid curve)
and $\Delta_\alpha=-0.4$ (dashed curve).
}
\label{fig:transm_pure}
\end{figure}

Depending on the soliton parameter $\beta$ and the frequency detuning $\delta$ one
or several levels $k_z^{(j)}$ may enter the open channel band; $k_z^{(j)}\le n_{+,0}$.
In particular, for the case of coherent beams with $\delta=0$ the number of levels is fixed to $n=2$, and the first level $k_z^{(1)}$ enters the open channel band when $1<\beta/n_{0,0}\lessapprox 1.5$. However, in this case all the resonance effects
are suppressed due to the transparency of the NLS soliton\cite{ZS72}, see solid
black curve in Fig.~\ref{fig:transm_pure}. Introducing a frequency detuning
$\delta\ne 0$ between the probe and the soliton beams, we break the soliton
transparency so that we may expect a Fano resonance in certain parameter regimes.
To confirm this, we have computed the transmission coefficient $T(k_x)$ for a
detuned probe beam using the simple ansatz 
\begin{eqnarray}
\label{npms}
n_{\pm,0}^2=n_{0,0}^2 \cdot (1\pm\Delta_n)^2\;,\\
\label{apms}
\alpha_{\pm}=\alpha_0 \cdot (1\pm\Delta_\alpha)^2\;
\end{eqnarray}
for the coefficients $n_{\pm,0}$ and $\alpha_{\pm}$ entering Eqs.~(\ref{2ch-1})
and (\ref{2ch-2}) for the two channel amplitudes\footnote{The correct estimate
of $n_{\pm,0}$ and $\alpha_{\pm}$ in accordance with Eqs.~(\ref{npm}) and
(\ref{apm}) involves $\omega$ dependencies $n(\omega)$ and $\alpha(\omega)$,
specific for the material of the waveguide and the frequency range.}.
The results shown in Fig.~\ref{fig:transm_pure} demonstrate the appearance of
single or even multiple Fano resonances -- resonant total reflection -- at certain
$k_x$ values of the probe beam. The resonance position is strongly depending both
on the soliton parameter $\beta$ and on the frequency detuning $\delta$ between
the soliton and the probe beams. It is also important, that the resonances shown
in Fig.~\ref{fig:transm_pure} occur for large enough values of $\Delta_n$ and 
$\Delta_\alpha$. Therefore, when trying to experimentally observe Fano
resonances it is highly desirable to operate in those frequency domains, where
the dependencies $n(\omega)$ and $\alpha(\omega)$ (which determine the differences
between $n_+$ and $n_-$, $\alpha_+$ and $\alpha_-$, respectively) are strong enough.

\begin{figure}
\begin{tabular}{cc}
\includegraphics[angle=270, width=0.45\textwidth]{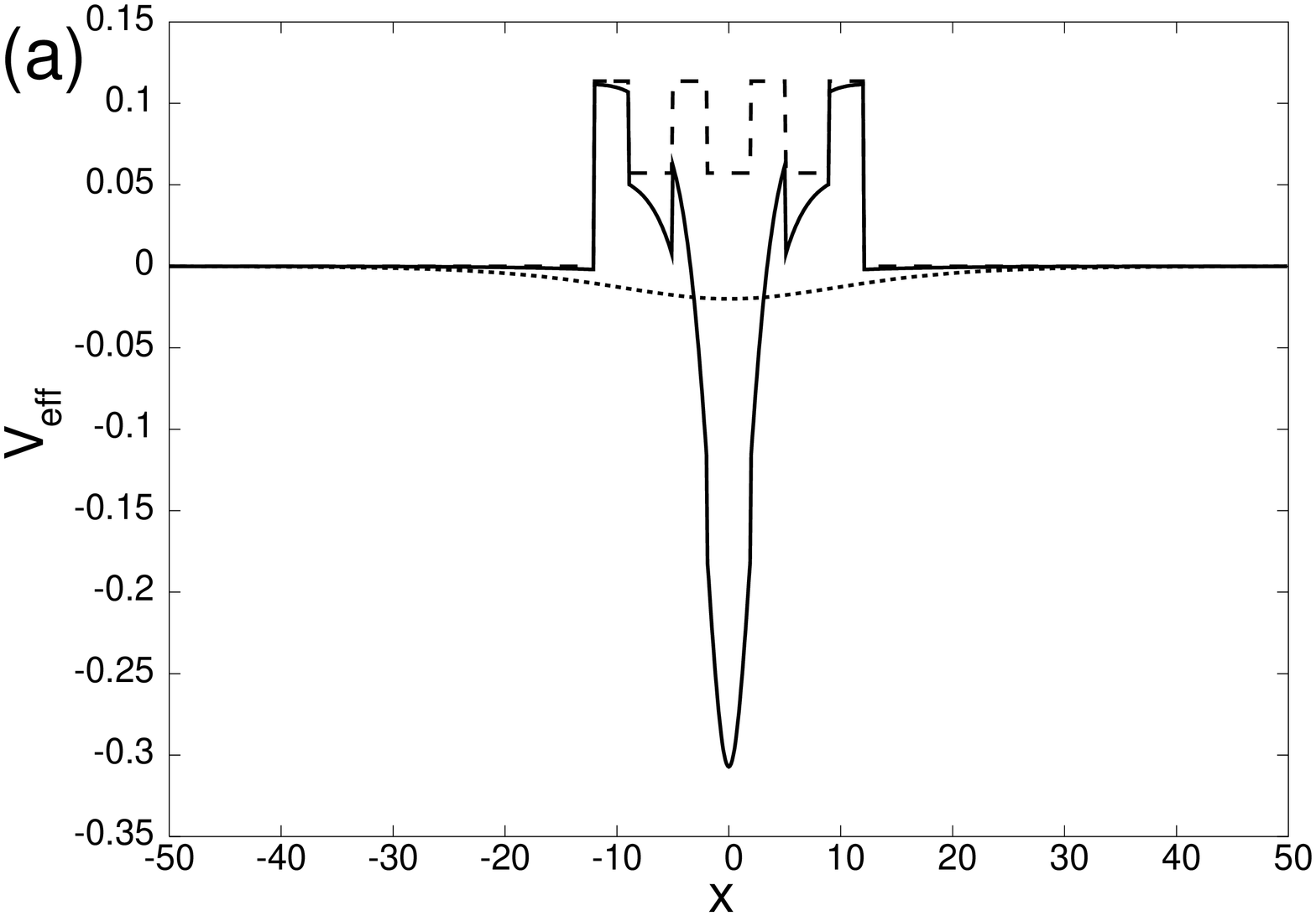}&
\includegraphics[angle=270, width=0.45\textwidth]{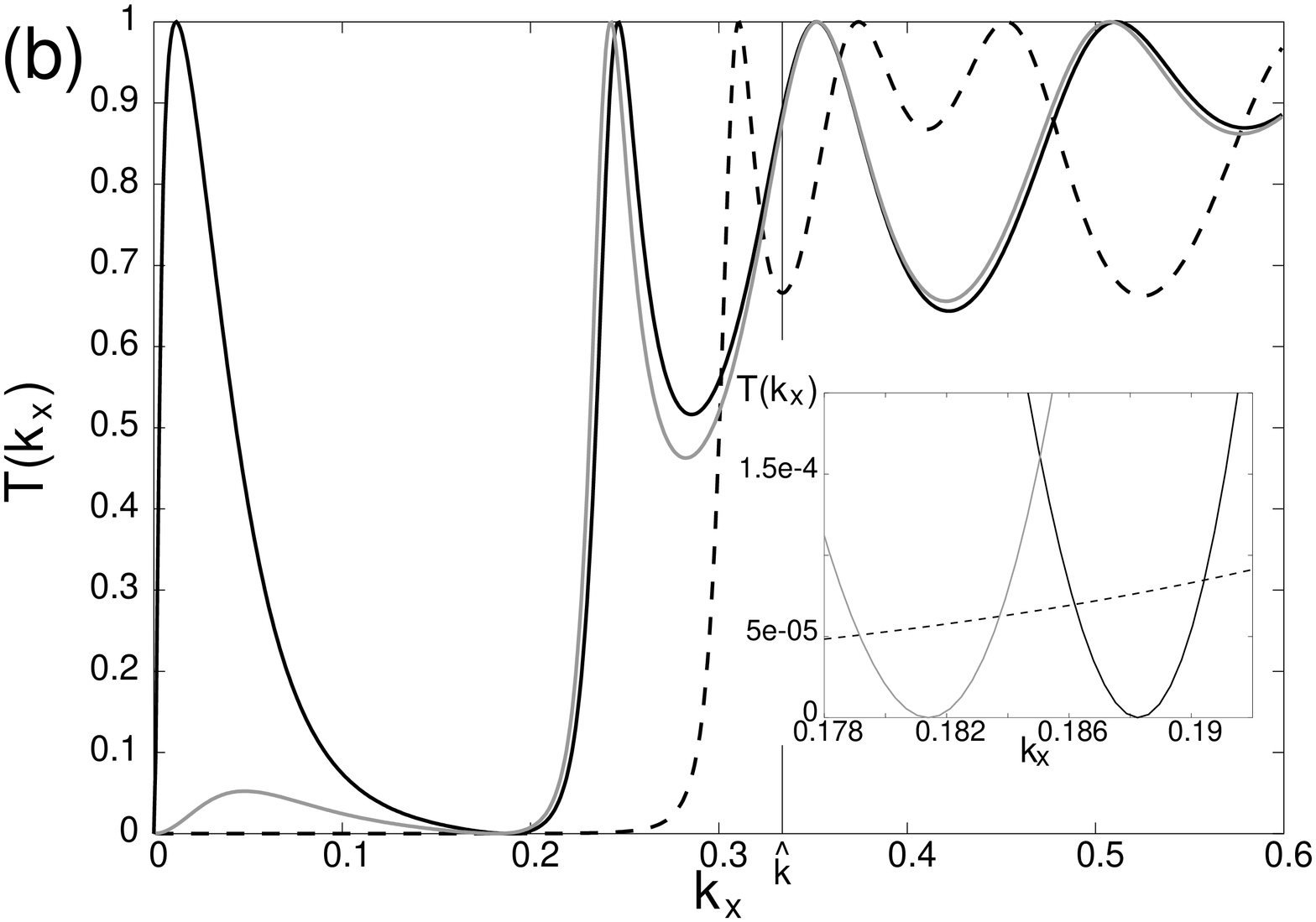}
\end{tabular}
\caption{(a) Effective potential $V_{eff}(x)$~(\ref{veff}) (solid line) and its 
geometrical part $n_{0,0}^2-n_0^2(x)$ (dashed line) for the TW configuration and
parameters: $n_{0,0}=1.44$, $n_{0,1}=1.4$, $n_{0,2}=1.42$, $L_s=12$,
$L_b=4$, $\beta \approx 1.4417$, $\beta^2-n_{0,0}^2=5.0\cdot 10^{-3}$, $\alpha_0=1.0$.
The dotted line shows the effective potential for the case of a homogeneous core $n_0(x)\equiv n_{0,0}=1.44$ with the same value of the soliton parameter
$\beta$; (b) Transmission coefficient $T(k_x)$ for the system without a soliton
(dashed line), and with solitons characterized by different parameters:
$\beta\approx 1.4417$, $\beta^2-n_{0,0}^2=5.0\cdot10^{-3}$ (solid black line),
$\beta \approx 1.4434$, $\beta^2-n_{0,0}^2=0.01$ (solid gray line). Other
model parameters are the same as in (a). The vertical line indicates the critical
value $\hat{k}\approx 0.337$ (see the text for details). Note, that $\delta=0$
here (coherent interaction).}
\label{fig:triplet}
\end{figure}

In order to observe Fano resonances for coherent beams with $\delta=0$, an
inhomogeneous refractive index $n_0(x)$ [which coincides with $n_{\pm}(x)$]
with the specially designed structure is needed \cite{FFG+05}. Here we discuss the 
so-called triple well (TW) configuration of the planar waveguide core, which was 
demonstrated in Ref.\cite{FFG+05} to support a Fano resonance under certain conditions. 
The refractive index $n_0(x)$ inside the core is locally decreased in the vicinity
of the scattering center where the optical soliton is formed, $n_0(x)=n_{0,1}<n_{0,0}, |x|<L_s$. In addition, in order to stabilize the soliton, local wells with a higher 
value of the refractive index $n=n_{0,2}$ ($n_{0,1}<n_{0,2}<n_{0,0}$) are introduced 
inside the $n=n_{0,1}$ section, the width of each well is $L_b<L_s$. The corresponding 
structure of the effective potential $V_{eff}(x)$~(\ref{veff}) is shown in 
Fig.~\ref{fig:triplet}(a), and the transmission coefficient computed for the TW
configuration both without and with solitons with different parameters is plotted in 
Fig.~\ref{fig:triplet}(b).

Note, that the $T(k_x)$ dependence for the TW configuration without a soliton
has several peculiarities. First of all, there is always a critical value
$\hat{k}$, below which the transmission coefficient is rather small. This fact has
a direct connection to the well known effect of the 'total internal reflection'
at the boundary between the homogeneous core and the central section with a
\emph{lower} value of the refractive index. Additionally, we observe several
transmission peaks at larger $k_x$ values connected with internal modes and
determined by the specific internal configuration of the TW section, i.e. by
the $n_0(x)$ dependence\cite{SWM93}.

The optical soliton has a twofold effect on the $T(k_x)$ dependence.
The DC part of the scattering potential is modified (see Fig.~\ref{fig:triplet}),
which leads to a complete restructuring of all the internal modes, and therefore
all the resonant transmission peaks are shifted, as seen in Fig.~\ref{fig:triplet}(b).
Most importantly, the AC part of the scattering potential becomes active. We 
observe a resonant reflection ($T=0$) of the probe beam in the
'dark region' $k_x<\hat{k}$ [see inset in Fig.~\ref{fig:triplet}(b)], provided
the soliton intensity is not too high ($\beta\lesssim 0.1$)\cite{FFG+05}. The
position of the Fano resonance is determined by the soliton parameter $\beta$.

The combination of the two above methods -- frequency detuning between the beams and modification of the linear refractive index $n_0(x)$ -- provides us with a
good flexibility for tuning the resonance parameters in the desired range of
the incident angles.

\section{Wavepacket scattering: spectral "hole burning"}
\label{sec_wp}

In order to investigate the possibility of experimental observation of the
above resonances, we perform an analysis of wavepacket scattering
by optical solitons. Let us focus on the particular case of coherent scattering,
when both the soliton and the probe beam have the same frequency $\omega_0$
($\delta=0$). We consider the above TW configuration with the parameters
as in Fig.~\ref{fig:triplet} and the soliton with $\beta\approx
1.4434$, $\beta^2-n_{0,0}^2=0.01$ [the corresponding transmission coefficient
for plane waves is shown in Fig.~\ref{fig:triplet}(b), gray curve].

To reveal possible nonlinear interaction effects, we will not use the
linearization procedure for the probe beam here. For this purpose it is
convenient to use the single envelope $E(x,z)$ which describes both beams.
Thus, instead of the ansatz~(\ref{three_beam}), we take
\begin{equation}
\mathcal{E}_{\bot}(x,z,t)=\frac12 \left \{E(x,z) \exp[i\beta z - i\omega_0 t] + 
c.c.\right\},
\label{single_beam}
\end{equation}
where we have separated the fast oscillating term $\exp[i\beta z]$ (i.e. we perform a
transformation to the rotating frame, in which the soliton has a $z$-independent profile). 
For the envelope $E(x,z)$ the following equation is obtained
\cite{FFG+05}:
\begin{equation}
\label{NNLS}
i2 \beta \frac{\partial E}{\partial z} + \frac{\partial^2 E}{\partial z^2}+ 
\frac{\partial^2 E}{\partial x^2} + \left[ n_0^2(x)-\beta^2\right] E+
\alpha_0 \left|E\right|^2 E =0\;.
\end{equation}
Considering small incident angles of the probe beam, we have $\partial E/\partial
z \ll \beta E$, and the second term in Eq.~(\ref{NNLS}) can be dropped\footnote{We
use this approximation mainly for technical reasons, in order to simplify the numerical 
integration of Eq.~\ref{NNLS}. Treating the full problem, one needs to use special 
techniques developed for non-paraxial beams\cite{CMN01}.}. This corresponds to the 
so-called \emph{paraxial approximation}, widely used in literature\cite{kivshar,akhmediev}.
In terms of the probe beam parameter $k_x$ the condition of validity of the
paraxial approximation is $(\beta-\sqrt{n_{0,0}^2-k_x^2})/\beta \ll 1$.
For our model and soliton parameter values this condition
is fullfiled for $|k_x|\lesssim 0.6$, i.e. essentially within the whole range
of $k_x$ presented in Fig.~\ref{fig:triplet}(b).
\begin{figure}
\begin{tabular}[t]{cc}
\includegraphics[angle=270, width=0.4\textwidth]{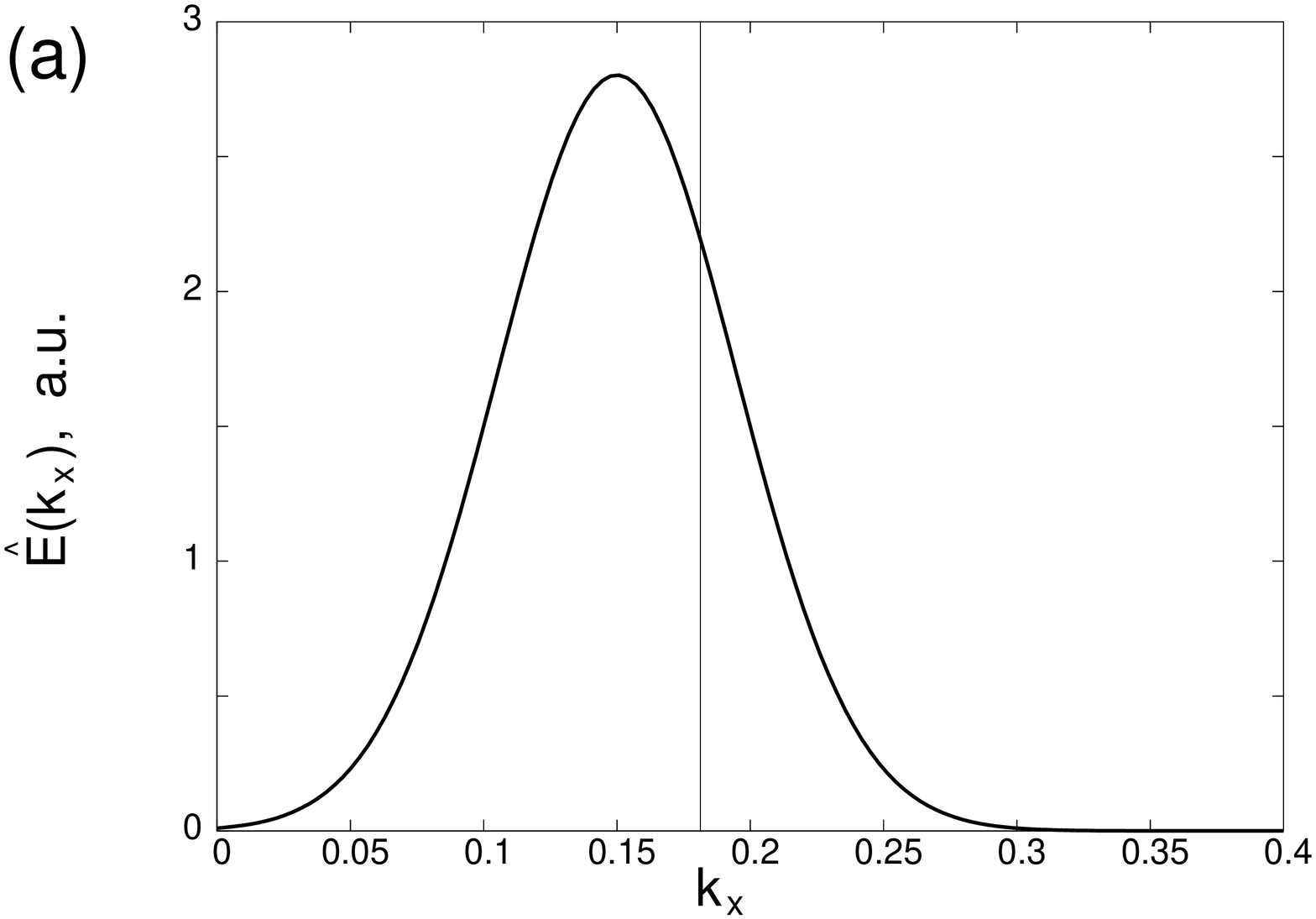}&
\includegraphics[angle=270, width=0.48\textwidth]{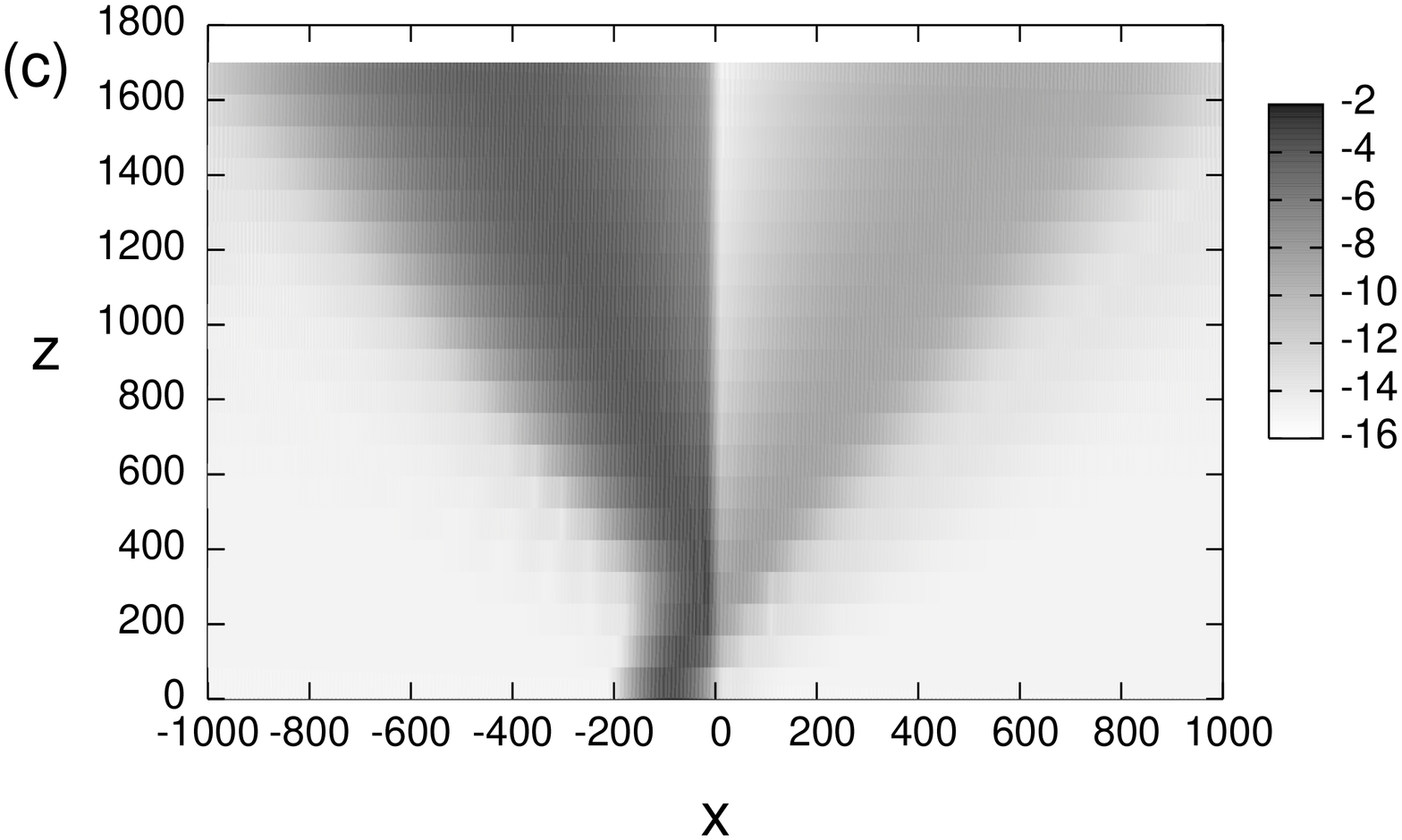}
\\
\includegraphics[angle=270, width=0.4\textwidth]{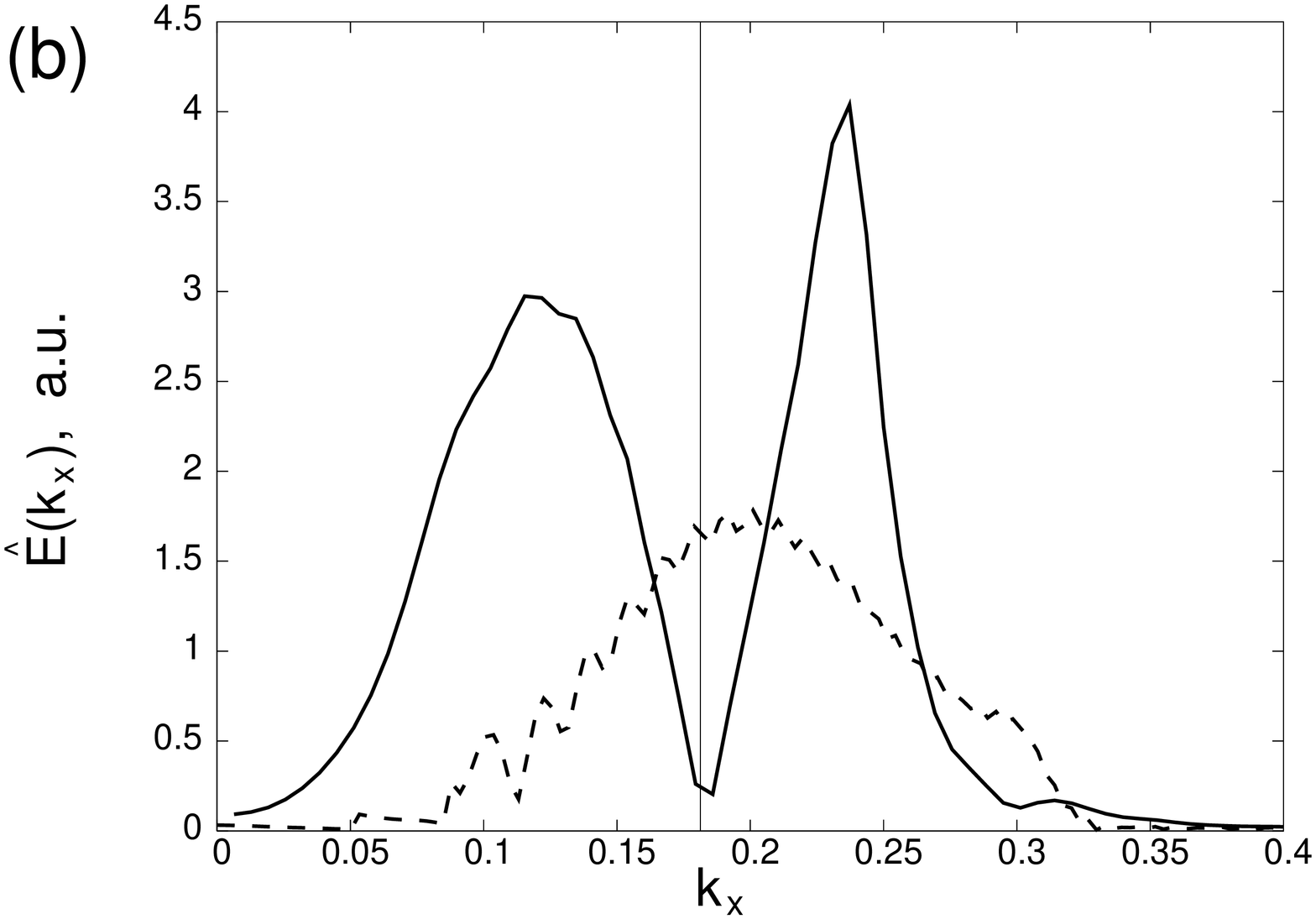}&
\includegraphics[angle=270, width=0.48\textwidth]{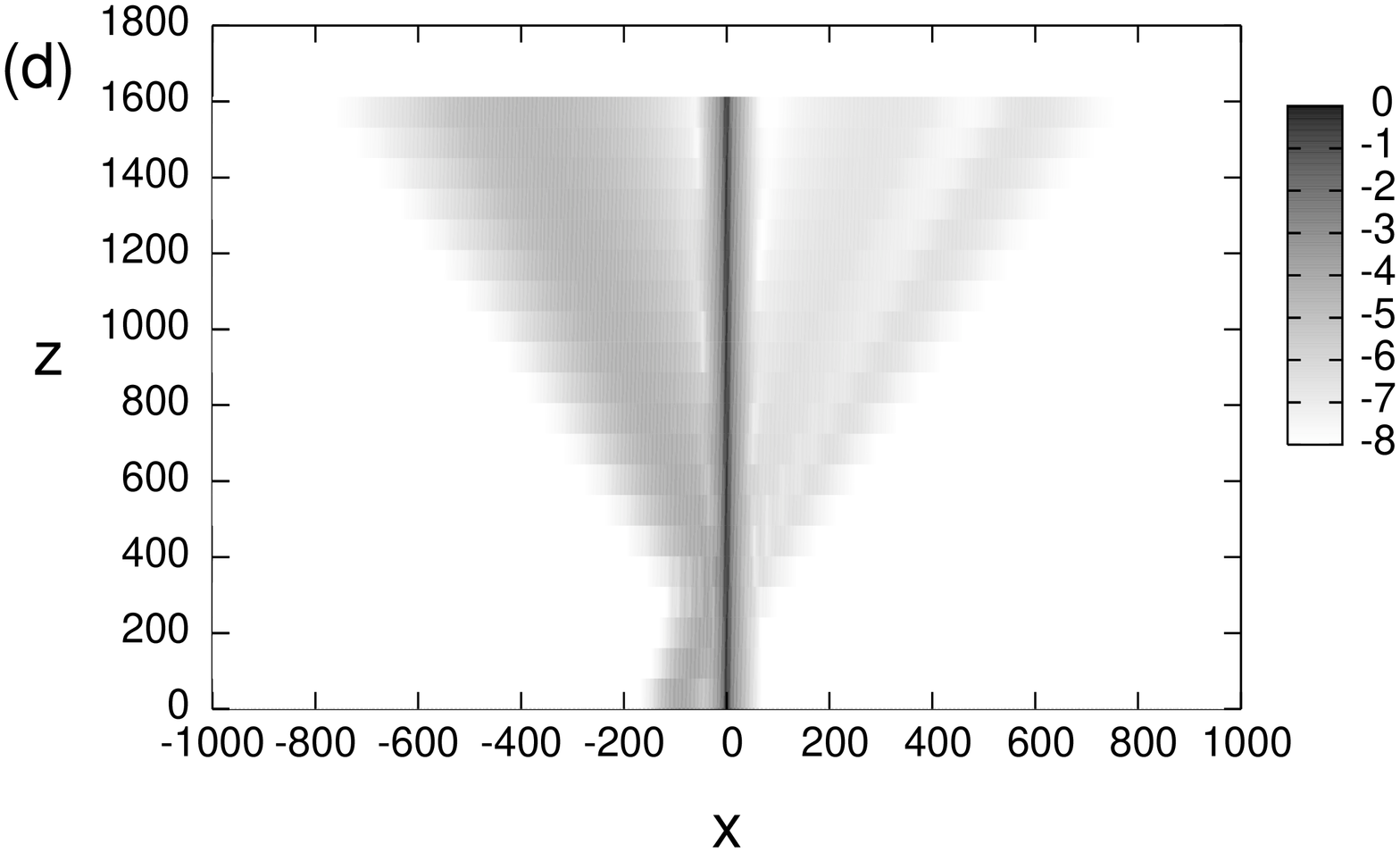}
\end{tabular}
\caption{(a),(b) Spatial Fourier transforms of the initial and transmitted wavepackets.
Vertical lines indicate the predicted resonance position. Dashed line in (b) corresponds
to the scattering process without the soliton. (c),(d) Density plots of the light power
distribution  $\log(|E(x,z)|^2)$ within the slab in the process of probe light scattering 
without and with the soliton, respectively.
\label{fig:wp}
}
\end{figure}

Treating $z$ as an effective time, we solve eqs.~(\ref{l+OB}) and (\ref{r+OB})
numerically as an initial value problem. In the input facet $z=0$ we add a small-amplitude Gaussian wavepacket to the soliton $C(x)$,
\begin{eqnarray}
E(x,z=0)=C(x)+A\exp\left[-\delta(x-x_0)^2\right]\exp[ik_0 x]\;.
\end{eqnarray}
We take $A=0.01,\;\delta=10^{-4},\;k_0=0.15$.  The wavepacket is considerably
broad in $k_x$-space and centered near the predicted above position of the
Fano resonance $k_x\approx 0.181$, see Fig.~\ref{fig:wp}(a).

The resulting light power distribution $|E(x,z)|^2$ in the slab is shown in
Figs.~\ref{fig:wp}(c) and (d) for the system without and with the soliton,
respectively. In the former case the scattering process is performed purely
by geometrical part of the effective potential~(\ref{veff}). Apparently, the
system with the soliton is more transparent for the wavepacket [note the different
scales in Figs.~\ref{fig:wp} (c) and (d)]. This result is in full agreement with
the computed transmission coefficient for plane waves, see
Fig.~\ref{fig:triplet}(b). Indeed, apart from the very narrow region of $k_x$ around
the resonance position $k_x\approx 0.181$, the soliton-induced nonlinear correction
to the DC part of the effective scattering potential~(\ref{veff}) [see Fig.~\ref{fig:triplet}(a)] makes the system more transparent at low values of
$k_x$. Thus, in order to observe the Fano resonance in terms of the transmission 
coefficient, one needs to operate with extremely narrow in $k$-space wavepackets
(with the effective width of the order of $\Delta k\lesssim 10^{-3}$). This
may introduce certain technical problems.
However, the resonance is much more pronounced in Fourier space and can be
observed with wide wavepackets as well: it manifests itself by filtering
out the resonant component from the initial wavepacket, see
Fig.~\ref{fig:triplet}(b). In real space it results in the wavepacket splitting,
clearly observed in the transmitted light area, see Fig.~\ref{fig:triplet}(d).
We believe, that such a spectral hole burning effect opens a promising way
for an experimental detection of the Fano resonance within the described setup.

\section{Conclusions}
\label{sec_concl}
To conclude, we propose experimental setups for a direct observation
of Fano resonances in the scattering of light by optical solitons. Such experiments
would be of a great importance, as they could confirm several theoretical
predictions of resonance phenomena in plane wave scattering by localized
nonlinear excitations\cite{FMF03,FMF+03,FFG+05}. The scattering process is
performed in a planar waveguide core. Two possible ways to obtain resonances in
the transmission are either to introduce a frequency detuning between the probe
and the soliton beams, or to insert a specially designed inhomogeneous refractive
index section inside the core. In both cases all the resonance effects, predicted
by our analysis, can be easily tuned into the experimental window,
as they strongly depend on the soliton intensity and the frequency detuning.
This could be also of importance from the practical point of view, giving an 
opportunity to use such resonance effects as tunable filters.

\begin{acknowledgments}
We would like to thank Shimshon Bar-Ad, Joachim Brand, Mordechai Segev and Lawrence Schulman 
for useful and stimulating discussions.
\end{acknowledgments}

\bibliography{jabbr,optics}  
\bibliographystyle{spiebib}

\end{document}